# Detection of atomic spin labels in a lipid bi-layer using a single-spin nanodiamond probe

Stefan Kaufmann[1,2*], David A. Simpson[1,3*], Liam T. Hall[1], Viktor Perunicic[1], Philipp Senn[4,5],

Steffen Steinert[6], Liam P. McGuinness[1], Brett C. Johnson[7], Takeshi Ohshima[7], Frank Caruso[4],

Jörg Wrachtrup[6], Robert E. Scholten[8], Paul Mulvaney[2], Lloyd C. L. Hollenberg[1,3+]

[1] Centre for Quantum Computation and Communication Technology, School of Physics,

University of Melbourne, Victoria 3010, Australia.

[2] School of Chemistry and Bio21 Institute, University of Melbourne, Victoria 3010, Australia.

[3] Centre for Neural Engineering, University of Melbourne, Victoria 3010, Australia.

[4] Department of Chemical and Biomolecular Engineering, University of Melbourne, Victoria

3010, Australia.

[5] Bionics Institute, Victoria 3065, Australia.

[6] 3. Physikalisches Institut, Research Center SCOPE, and MPI for Solid State Research,

University of Stuttgart, Pfaffenwaldring 57, 70569 Stuttgart, Germany.

[7] Semiconductor Analysis and Radiation Effects Group, Japan Atomic Energy Agency, Takasaki,

Japan.

[8] Centre for Coherent X-Ray Science, School of Physics, University of Melbourne, Victoria

3010, Australia

*These authors contributed equally to this work.

[+]Corresponding author: Lloyd Hollenberg - School of Physics, University of Melbourne,

Victoria, 3010, Australia. Tel: +61 3 8344 4210, Email: lloydch@unimelb.edu.au

*Classification:* PHYSICAL SCIENCES: Applied Physical Sciences

**Abstract**

Magnetic field fluctuations arising from fundamental spins are ubiquitous in nanoscale biology, and are a rich source of information about the processes that generate them. However, the ability to detect the few spins involved without averaging over large ensembles has remained elusive. Here we demonstrate the detection of gadolinium spin labels in an artificial cell membrane under ambient conditions using a single-spin nanodiamond sensor. Changes in the spin relaxation time of the sensor located in the lipid bilayer were optically detected and found to be sensitive to near-individual (4 ± 2) proximal gadolinium atomic labels. The detection of such small numbers of spins in a model biological setting, with projected detection times of one second, opens a new pathway for in-situ nanoscale detection of dynamical processes in biology.

The development of sensitive and highly localized probes has driven advances in our understanding of the basic processes of life at increasingly smaller scales (1). In the last decade there has been a strong drive to expand the range of probes that can be used for studying biological systems (2-6), with emphasis on the detection of atoms and molecules in nanometer sized volumes in order to gain access to information that may be hidden in ensemble averaging. However, at present there are no nanoprobes suitable for directly sensing the weak magnetic fields arising from small numbers of fundamental spins in nanoscale biology, occurring naturally (e.g. free-radicals) or introduced (e.g. spin-labels), which can be a rich source of information about processes at the atomic and molecular level. Magnetic resonance techniques such as electron spin resonance (ESR) have played an important role in the development of our understanding of membranes, proteins and free radicals (7); however, ESR sensitivity and resolution are fundamentally limited to mesoscopic ensembles of at least $10^7$ spins with a sensitivity of ~ $2 \times 10^9$ spins/(Hz)$^{1/2}$ (8). In a typical ESR application small electron spin label moieties are attached to the system of interest and their environment is investigated through spin measurements on the labels. Because of the large ensemble required nanoscopic detail at the few spin level can be lost in the averaging process. Recently, magnetic resonance force microscopy techniques have demonstrated single spin detection (9-11), but these require cryogenic temperatures and vacuum. Here we demonstrate a nanoparticle probe – a nitrogen-vacancy spin in a nanodiamond – which is situated in the target structure itself and acts as a nanoscopic magnetic field detector under ambient conditions with non-contact optical readout. We employ this probe to detect near-individual spin labels in an artificial cell membrane at a projected sensitivity of ~ 5 spins/(Hz)$^{1/2}$, effectively bridging the gap between traditional ESR ensemble based techniques and the ultimate goal of few-spin nanoscale detection under biological conditions.

**Results and discussion**

The overall set-up of our experiment is shown schematically in Figure 1A. The nanoparticle probe is the single spin of a nitrogen-vacancy (NV) centre contained in a nanodiamond particle (Figure 1B). NV-nanodiamonds are attractive nanoprobes for biology as they exhibit low toxicity and are photostable (12-14). Here we investigate the magnetic field sensing capabilities of the NV centre in a biological context. Around these NV-nanodiamonds we formed a supported lipid bilayer (SLB) on a glass substrate. The SLB was created with lipids labeled with Gd$^{3+}$, a common MRI contrast agent with spin 7/2 (Figure 1C) which produce characteristic magnetic fluctuations in the lipid environment (Figure 1D) and form our detection target. The ground state of the NV centre has a zero-field splitting of $D = 2.87$ GHz between the $|0\rangle$ and $|\pm1\rangle$ spin levels and optical-based spin state readout is possible due to the significantly lower fluorescence of the $|\pm1\rangle$ compared to the $|0\rangle$ state (15) (Figure 1E). These properties have led to demonstrations of DC and AC magnetic field detection using single NV spins (16-18), and wide-field detection schemes employing ensembles of NV centres (19-22). For biological applications, where atomic level processes produce magnetic field fluctuations, it has been proposed that changes in the quantum decoherence of the NV spin could provide a more sensitive detection mechanism (23-25). Ensemble relaxation based imaging has been demonstrated with sub-cellular resolution and detection of ~$10^3$ spins (26). Towards the goal of using single NV spins as *in situ* nano-magnetometer probes in biology, quantum measurements on single NV spins have been carried out in a living cell (27), and with the recent detection of external spins in well controlled environments (28-30), the critical milestone is to demonstrate nanoscopic external spin detection in a biological context. Here we achieve this with near individual spin sensitivity by monitoring the relaxation time ($T_1$) of a single NV spin probe (Figure 1E, F) embedded in the SLB target itself. In addition, our results indicate that the NV spin is sensitive to cross-lipid magnetic fluctuations arising from the small number of Gd labeled lipids in the SLB in the vicinity of the nanodiamond (Figure 1A, D).

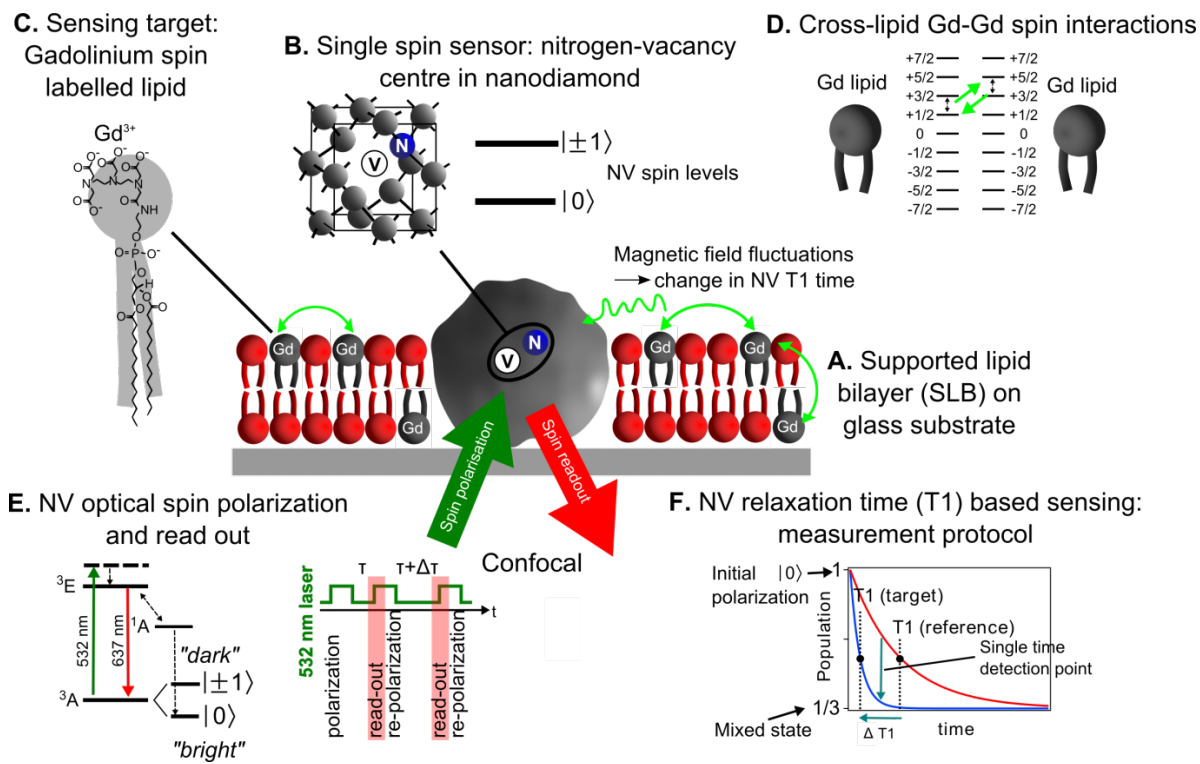

**Figure 1:** Schematic of nanoscopic detection of spin labels in an artificial cell membrane using a single-spin nanodiamond sensor. **A.** A supported lipid bilayer (SLB) is formed around a nanodiamond immobilized on a glass substrate. **B.** The nanodiamond contains a single nitrogen-vacancy (NV) optical centre which acts as a single spin sensor by virtue of the magnetic levels in the ground state. **C.** Gadolinium (Gd) spin labeled lipids are introduced into the SLB. **D.** Magnetic field fluctuations arising from Gd spin labels affect the quantum state of the NV spin, measured through the NV relaxation time, $T_1$. **E.** The electronic energy structure of the NV centre showing the fluorescent cycle and optical spin readout of the spin states $|0\rangle$ and $|\pm1\rangle$, and the protocol for the $T_1$ measurement. **F.** A schematic illustration of the $T_1$ measurement. The relaxation of the NV spin in the target environment is compared to that in the reference environment. Measurement at a single time point in the evolution allows faster detection.

The SLB was formed on the nanodiamond-glass substrate in tris-buffered saline (TBS) solution via the vesicle fusion technique (31) (Figure 2A), and characterized by both fluorescence recovery after photo-bleaching (FRAP) and atomic force microscopy (AFM) (Figure 2B, and Supplementary Information). AFM investigations showed that the SLB formed mainly around the nanodiamond particles (mean size = 15.9 ± 9.5 nm). The $T_1$ time of the NV spin was measured by optically polarizing into the $|0\rangle$ state and measuring the probability $P_0(t)$ of finding the NV in the initial $|0\rangle$ state at a later time $t$ through spin state fluorescence contrast. In a low background field the $|\pm1\rangle$ spin states of the NV are approximately degenerate so the $P_0(t)$ decays as:

$$P_0(t) = (1/3)(1 + \exp(-t/T_1) + \exp(-2t/T_1)). \tag{1}$$

This form is a consequence of the 3-state nature of the NV spin manifold and the broad spectrum of the target spin bath (see Supplementary Information for the derivation). Figure 3A (green dashed curve) shows $P_0(t)$ (normalised fluorescence) for a single NV centre (NV1) under the TBS control conditions. Fitting to Equation (1) we obtain a reference relaxation time $T_1$[TBS] = 117±11 μs. $T_1$[TBS] includes intrinsic nanodiamond sources of relaxation such as N donors (electron spin bath), and $^{13}$C spins (nuclear spin bath), but is dominated by surface electronic spins fluctuating in the GHz regime. After the formation of a SLB without Gd spin labels (under the same TBS conditions), the measurement was repeated on the same NV (green data points) to give $T_1$[SLB] = 139±18 μs, showing no statistically significant change for the unlabeled lipid case. Removal of the SLB, which included a 2 min oxygen plasma etching step, was followed by a new TBS reference $T_1$ measurement (blue dashed curve). Upon formation of the SLB with Gd spin labeled lipids at 10% (w/w) concentration (blue data points), we observed a 61% reduction in the NV relaxation time to $T_1$[SLB:Gd] = 48±3 μs.

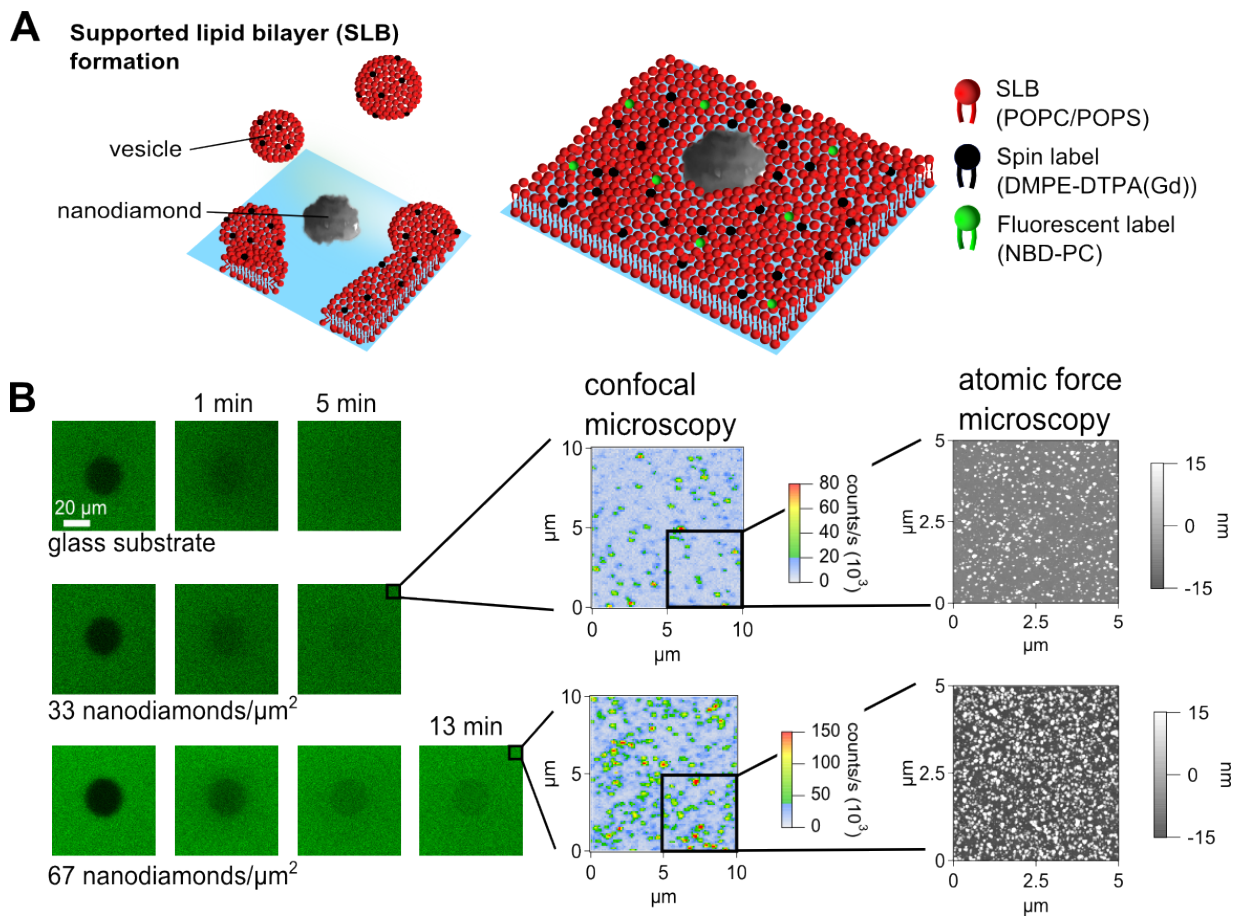

**Figure 2:** Formation and characterisation of a supported lipid layers (SLB) on nanodiamond/glass substrates. **A.** SLBs were formed by the vesicle fusion technique and characterized by fluorescence recovery after photobleaching (FRAP). **B.** Confocal images of FRAP measurements directly after the bleaching cycle for a SLB+10% (w/w) of the Gd spin label and 1% (w/w) of a fluorescent label. For characterisation, three different substrate conditions were investigated. Confocal and atomic force microscopy images show nanodiamond density and size distribution.

To demonstrate the consistency of the effect, this sequence was performed for five NV centres (NV1-NV5), in distinct nanodiamonds. Figure 3B shows the percentage changes in the $T_1$ times for all five NV centres relative to their TBS reference values. For the SLB with no spin labels there is no statistically significant change verifying that the NV $T_1$ relaxation time remains intact under control conditions. For the SLB labeled with 10% Gd the average reduction in the NV $T_1$ relaxation time from the reference value is 74 ± 6 %, and is remarkably consistent across the set of nanodiamonds. We next investigated the Gd concentration dependence of the change in the $T_1$ relaxation time. In Figure 3C the $T_1$ curves of a single NV centre (NV6) are given for 10% and 1% Gd concentrations, showing the change in $T_1$ decreasing as the number of proximate Gd spins is reduced. The percentage changes in $T_1$ from the respective TBS reference values as a function of Gd concentration are given in Figure 3D for another set of four individual NV centres (NV6-NV9) confirming this trend.

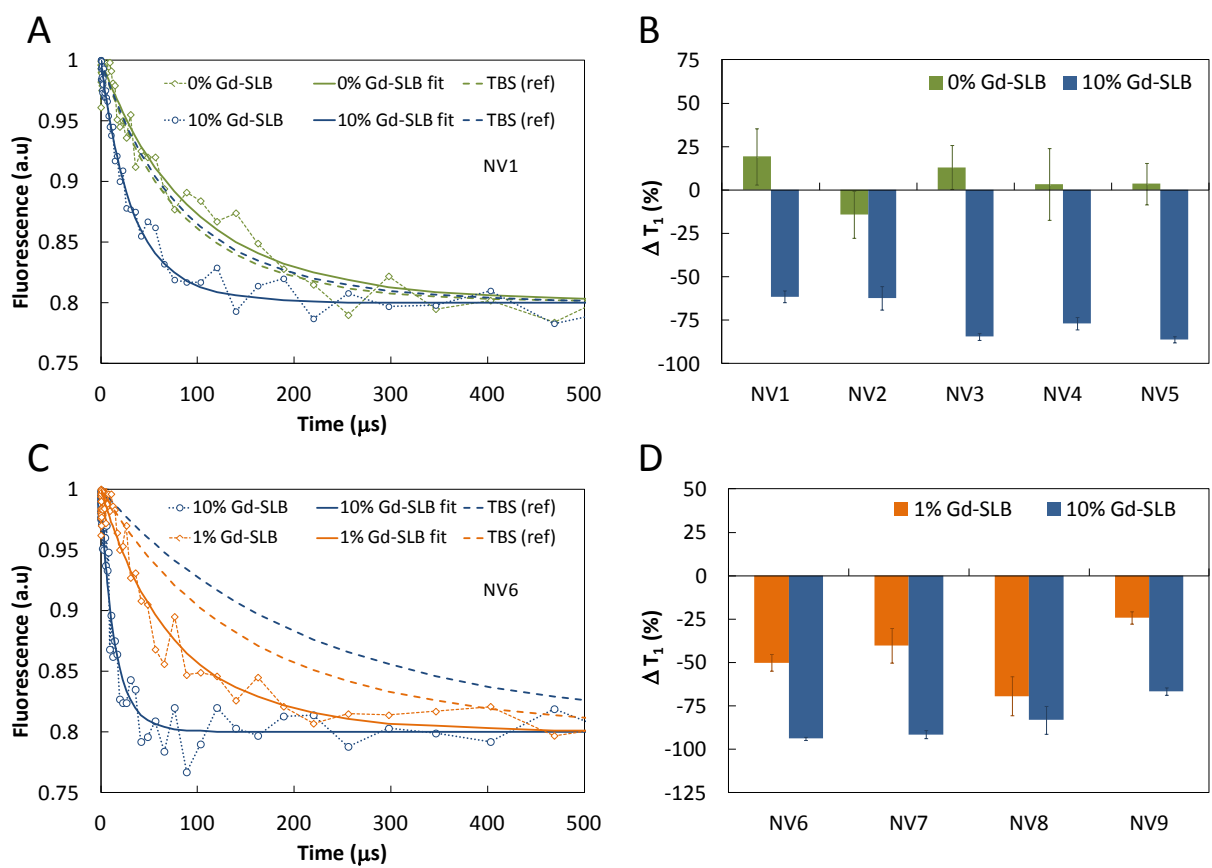

**Figure 3:** Detection of spin labeled lipids in a supported lipid bilayer (SLB) using the $T_1$ time of single NV spins in nanodiamonds. **A.** Relaxation of the spin of a single NV centre (NV1) in a nanodiamond in a SLB without spin labels (green) and in a SLB with 10% (w/w) Gd spin labels (blue). **B.** The percentage change in $T_1$ (relative to TBS) of five single NV spins (NV1-NV5) in distinct nanodiamonds: SLB+0% Gd (green) and SLB+10% Gd (blue). **C.** Relaxation of the spin of NV6 for SLB+ 10% Gd (blue) and SLB+1% Gd (orange). **D.** Concentration dependence of the percentage change in $T_1$ for a set of centres, NV6-NV9. The data in B and C, after fitting to determine the relaxation times, have been scaled to the same asymptotic value for ease of presentation. All error bars represent the fitting uncertainties at the 95% confidence interval. Solid curves are the corresponding fits of the data to Equation (1) which determine $T_1$. Dashed curves correspond to the fitted data of the corresponding reference TBS measurement prior to each SLB measurement.

In order to understand these results quantitatively we consider the quantum evolution of the NV spin in the presence of magnetic field fluctuations arising from the various atomic processes involving the Gd spins in the SLB (see Supplementary Information). The characteristic timescales of these fluctuations produce specific changes in the measured $T_1$ relaxation time of the NV spin from the reference value. In terms of external sources of decoherence, the quantum evolution of the NV spin is determined by the overall spectral distribution function, $S_{f_e}(\omega)$, which is governed by the characteristic environmental magnetic field fluctuation frequency, $f_e$. The Gd-labeled lipid environment produces magnetic field fluctuations due to cross-lipid Gd-Gd spin dipole interactions, motional diffusion of individual Gd-labeled lipids, and intrinsic Gd spin relaxation effects, each with characteristic frequencies $f_{\text{dip}}$, $f_{\text{dif}}$, and $f_{\text{in}}$ respectively that contribute to $f_e$. From $S_{f_e}(\omega)$ we obtain the NV relaxation time $T_1[\text{Gd}]$ due to Gd-lipid effects: $T_1[\text{Gd}] = (f_e^2 + (D - 2D_{Gd})^2)/2f_e B_{\text{eff}}^2$, where $B_{\text{eff}} = \sqrt{63\pi\sigma}(\mu_0 \gamma_{\text{NV}} \gamma_{\text{Gd}})/128 h^2$ is the characteristic Gd-NV dipolar magnetic field interaction, $h$ is the NV depth below the nanodiamond surface, $\sigma$ is the areal Gd spin density, and $D_{\text{Gd}}$ is the zero field splitting parameter of the Gd spin, with the dominant contribution to $T_1$ coming from the $2D_{\text{Gd}}$ transition. We note that the $4D_{\text{Gd}}$ and $6D_{\text{Gd}}$ transitions are too far detuned from D to have any significant effect on $T_1$. We obtain $f_{\text{dip}}$ by integrating the Gd-Gd dipolar auto-correlation function over a planar SLB distribution to give $f_{\text{dip}} = 63\mu_0 \hbar \gamma_{\text{Gd}}^2 (\pi\sigma)^{3/2}/32\pi$. For the diffusion of Gd-lipids in the SLB we obtain $f_{\text{diff}} = 9D_l/16h^2$ where $D_l$ is the Gd-lipid diffusion constant. The intrinsic Gd spin relaxation is characterized by $f_{\text{in}} \approx 1\,\text{GHz}$ (32). For the NV depths expected in these 16 nm nanodiamonds and Gd concentrations employed here, the effect of Gd-Gd interactions on the NV relaxation time dominate over Gd diffusion effects. Expanding around the low Gd density limit we arrive at a theoretical model for the NV relaxation time $T_1[\text{Gd}]$, due to proximate Gd labeled lipids:

$$T_1[\text{Gd}] = \frac{h^4}{w}(aw^{3/2} + b), \tag{2}$$

where $w$ is the %w/w Gd-lipid concentration, and $a$ and $b$ are constants involving the physical parameters $D$, $D_{\text{Gd}}$ and $f_{\text{in}}$ (see Supplementary Information). Because the nanodiamond surface spins comprise a two-dimensional distribution we also obtain $T_1[\text{TBS}] \propto h^4$. Hence there is a cancelation of the depth dependence in the percentage change in relaxation times, $\Delta T_1[\text{SLB:Gd}]/T_1[\text{TBS}]$, which may explain the uniformity of the effect across the set of NV-nanodiamonds shown in Figure 3B.

To directly compare our data with Equation (2) we extract $T_1[\text{Gd}]$ by subtracting the reference rate (i.e. $\Gamma_1[\text{Gd}] = \Gamma_1[\text{SLB:Gd}] - \Gamma_1[\text{TBS}]$) and in Figure 4 plot this quantity for all NV1-NV9 against Gd concentration, $w$. Generally, the data points lie in a NV depth range consistent with the measured AFM nanodiamond size distribution indicating that we observe $T_1[\text{Gd}]$ times in broad agreement with those predicted by Equation (2). The trend to shallower NV depths as we move from 10% → 1% Gd-lipid concentration is consistent with the etching step in the processing of the sample between the 10% and 1% measurements which is likely to have removed several nm of material from the nanodiamonds. However, we note that for low concentrations we also expect some deviation to the scaling in Equation (2) due to the statistics of low Gd spin numbers.

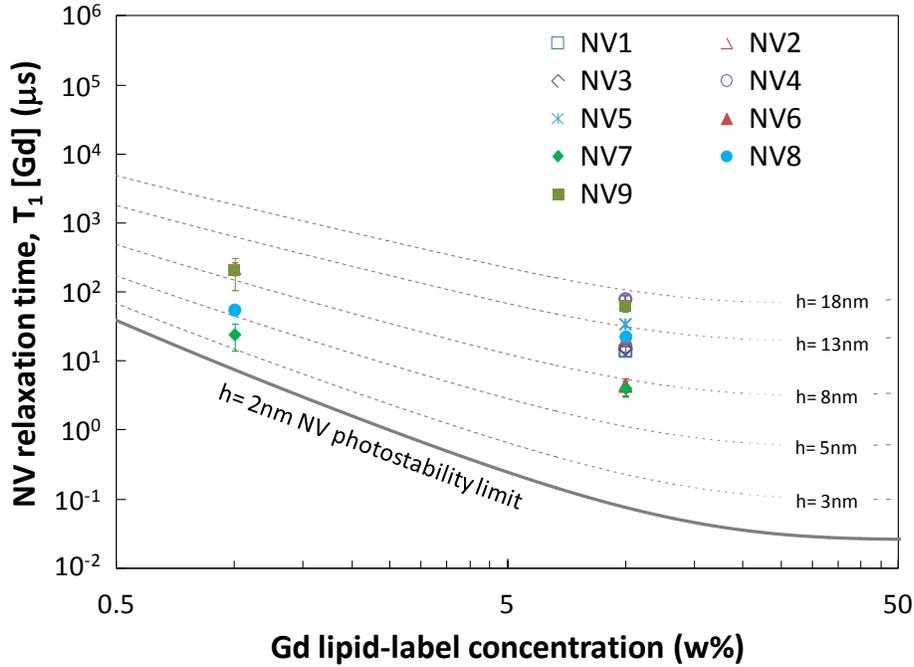

**Figure 4:** The NV relaxation time due to the presence of Gd spin labeled lipids, $T_1[Gd]$, as a function of Gd lipid concentration (%w/w) for all measured centres, NV1-NV9 (data points). Theoretical curves (dashed) based on Equation (2) are given for a range of NV depths, $h$, with a lower bound $h = 2$nm corresponding to the photostability limit (33).

The effective number of spins detected, $N_{eff}$, for a given Gd concentration $w$ is estimated by comparing the RMS field expectation $\langle B^2 \rangle^{1/2}$ at the NV contributing to the change in relaxation time integrated over all Gd spins in the membrane, to that of a single Gd spin at distance $h$, to obtain $N_{eff} \sim (3/8)\pi\sigma h^2$ (see Supplementary Information). Using the NV depth range $h \approx 8 \pm 5$ nm from the AFM distribution we arrive at a lower bound estimate of the effective number of spins detected of $4 \pm 2$ (1% Gd) and $28 \pm 24$ (10% Gd). Finally, from the data we can determine projected single time-point detection times assuming the reference TBS value has been well characterised and the measured fluorescence contrast is normalised. For the case of NV7, which had a relatively fast value of $T_1$[TBS], to detect the fluorescence change at the 95% CL corresponding to 1% Gd-SLB at a single time point requires a total measurement time of ~1.3s (see Supplementary Information).

In conclusion, we have demonstrated nanoscopic magnetic detection of spin labels in an artificial cell membrane using the $T_1$ relaxation time of a single spin NV-nanodiamond probe. Our results for Gd-labeled lipid concentrations down to 1% w/w correspond to the detection of near-individual Gd spin labels, with projected single time point detection times of order 1s. The data are in broad agreement with cross-lipid Gd-Gd spin interactions as the dominant atomic process detected. These results highlight the potential of the NV-nanodiamond system as a nanoscopic magnetic probe in biology which circumvents the fundamental problems associated with ensemble averaging.

**Materials and Methods**

*Nanodiamond preparation*

Nanodiamond were purchased from VanMoppes, Switzerland (SYP 0 - 0.03) and irradiated with high-energy electrons (2 MeV with a fluence of $1 \times 10^{18}$ electrons/cm$^2$) and vacuum annealed at 800 $^o$C for 2 hours at the Japan Atomic Energy Agency. The powder was heated to 425 $^o$C in air for 3 hours, dispersed in water and sonicated with a high-power sonicator (Sonicator 4000, Qsonica2) for 30 hours. The suspension was centrifuged at 12000 rcf for 120 s and the supernatant was removed and used as stock suspension. For the adsorption of nanodiamonds, cleaned glass substrates (No. 1, Menzel) were immersed in a 1 mg/ml solution of polyallylamine hydrochloride (PAH, 70kDa) in 0.5 M NaCl in Milli-Q water for 5 minutes. After rinsing with Milli-Q water nanodiamonds were adsorbed in the desired concentration in Milli-Q water for 5 minutes. The glass substrates were treated in a 40 W oxygen plasma (25% oxygen in argon, 40 sccm flow rate) for 5 minutes to remove the polyelectrolyte from the substrate. Atomic force microscopy (AFM) in air was performed with an Asylum MFP-3D in tapping mode with cantilevers from Olympus (AC160TS, 42 N/m).

*Formation of supported lipid bilayers (SLBs)*

All chemicals are of analytical grade reagent quality if not otherwise stated. The lipids (Avanti Polar Lipids Inc., USA) 1-palmitoyl-2-oleoyl-sn-glycero-3-phosphocholine (POPC), 1-palmitoyl-2-oleoyl-sn-glycero-3-phospho-L-serine (sodium salt) (POPS) and 1-palmitoyl-2-12-[(7-nitro-2-1,3-benzoxadiazol-4-yl)amino]dodecanoyl-sn-glycero-3-phosphocholine (NBD-PC) were dissolved in chloroform and the spin label 1,2-dimyristoyl-sn-glycero-3-phosphoethanolamine-N-diethylenetriaminepentaacetic acid (gadolinium salt) (DMPE-DTPA(Gd)) was dissolved in the mixture chloroform:methanol:water (65:35:8) (v:v). The lipids were mixed in the desired ratios and were dried under a stream of nitrogen for 1 hour followed by re-suspension in TBS buffer. TBS buffer was prepared by mixing 10 mM Tris(hydroxymethyl)-aminomethane (Sigma-Aldrich Pty. Ltd., Australia) and 150 mM sodium chloride (Sigma-Aldrich Pty. Ltd., Australia) in water with the pH adjusted to 7.4 by stepwise addition of 1 M hydrochloric acid. Milli-Q water (18.2 MΩ and at most 4 ppm (TOC), Millipore, USA) was used throughout. The buffer was filtered through 0.2 μm filters (PALL Acrodisc Syringe Filter) before use. Supported lipid bilayers (SLB) were formed with vesicles extruded 31 times through two stacked, 100 nm pore-size polycarbonate membranes (Avanti Polar Lipids Inc., USA) at a concentration of 0.1 mg/ml with an exposure time of 1 hour (overnight exposure for quantum measurements in liquid cell). The vesicle size was measured by dynamic light scattering resulting in an average size of 113 $\pm$ 3 nm (s.d). Before the formation of SLBs, 3 mM $Ca^{2+}$ ($CaCl_2$, Chem-Supply Pty Ltd, Australia) was added to the buffer solution. This improved the adhesion onto the negatively charged, glass surface (34). After the measurements the liquid cell was cleaned with 2% (v/w) of sodium dodecyl sulfate in water and isopropyl alcohol and thoroughly rinsed with water. Before quantum measurements the coverslip was treated in a 40 W oxygen plasma (25% oxygen in argon, 40 sccm flow rate) for 2 minutes.

*$T_1$ measurements*

Confocal imaging was performed on a home-built microscope using an oil-immersion objective (100x, Nikon). $T_1$ measurements were performed using a Spincore PulseBlaster card and a Fast ComTech P7889 Multiple-Event Time Digitizer card. Typical measurement time for a full evolution of a $T_1$ curve was 1 hour. Spin relaxation curves were fitted using Equation (1) in the main manuscript.

**Acknowledgements**

This work was supported by the Australian Research Council under the Centre of Excellence (CE110001027) scheme, the Centre for Neural Engineering, ERC Project SQUTEC, EU Project DINAMO, and the Baden-Wuerttemberg Foundation. SK was supported by a Swiss National Science Foundation fellowship PBEZP3-133244. FC is supported by the ARC Australian Laureate Fellowship scheme (FL120100030).